# Magnetotransport and thermal properties characterization of 55 K superconductor SmFeAsO$_{0.85}$F$_{0.15}$

Amit Srivastava[1], Anand Pal[2], Saurabh Singh[1], C. Shekhar[3], H.K. Singh[2], V.P.S. Awana[2,*] and O.N. Srivastava[1]

[1] Nano Science and Nanotechnology Centre, Department of Physics, Banaras Hindu University, Varanasi, U.P. 221005, India
[2] Quantum Phenomena and Application Division, National Physical Laboratory (CSIR) Dr. K. S. Krishnan Road, New Delhi-110012, India
[3] Max Plank Institute for Chemical Physics of Solids, Dresden-01187 Germany

## Abstract

. Correlation between structural/microstructural and magneto-thermal transport properties of FeAs-based SmFeAsO and SmFeAsO$_{0.85}$F$_{0.15}$ has been studied in detail. The Rietveld analysis of room temperature powder X-ray diffraction (*XRD*) data reveals that both the samples are single phase, with very small amount of rare earth impurity in the F doped compound. Electron microscopic investigations show that compounds have layered morphology and structure, with the individual grains being surrounded by amorphous layers. The average grain boundary thickness is ~5 nm. The F free material is found to be magnetic and shows the appearance of Fe spin density wave (*SDW*) like order at T < 150 K. The F doped compound (SmFeAsO$_{0.85}$F$_{0.15}$) shows the occurrence of superconductivity at T$_c$($\rho$=0, H=0)≈55 K, which decreases to ≈ 42 K at magnetic field (H) of 13 kOe. The superconducting transition was also confirmed by DC magnetization and AC susceptibility measurements. The intra-grain critical current density ($J_c$) calculated using the Bean critical state model is found to be around ≈5.26 ×10$^4$ A/cm$^2$ at 5 K in zero field (H=0). The dependence of thermally activated flux flow energy (U/k$_B$) on the applied magnetic field has been observed. *AC* susceptibility measurements at different amplitude of applied *AC* drive field confirm the granular nature of the superconducting compound. is confirmed. Both Fe (*SDW*) at 150K for SmFeAsO and 55K superconductivity in case of SmFeAsO$_{0.85}$F$_{0.15}$ sample has confirmed by Specific heat [$C_p(T)$] measurement too. Further Sm orders anti-ferro-magnetically at 4.5K for non-superconducting and at 3.5K for superconducting samples, also the entropy change is reduced significantly for the later than the former. Summarily complete physical property characterization for both non-superconducting SmFeAsO and 55K superconductor SmFeAsO$_{0.85}$F$_{0.15}$ samples is provided and discussed in the current article.

PACS number(s): 74.10.+v, 74.70.-b, 74.70. Dd, 74.25.F-

Key Words: Oxy-Arsenide superconductor family, Superconductivity, X-ray diffraction, Rietveld analysis, Magnetization and Thermally activated flux flow.

*Corresponding author's email: awana@mail.nplindia.ernet.in,

Fax No. +91-11-45609310: Phone no. +91-11-45609357

Web page: www.freewebs.com/vpsawana/

**Introduction**

The discovery of the new class of oxypnictide superconductors including iron-based LaFeAsO$_{1-x}$F$_x$ with critical temperature $T_c$ at 26 K [1] has provided new impetus to research in the area of high-temperature superconductivity, and has resulted in unravelling of several new issues in the domain of superconducting materials incorporating Fe-As layers as the fundamental structural unit [2-8]. Such newfangled materials have the general formula REFeAsO, where RE is a rare earth element. The structural unit consists of alternating RE-O and Fe-As layers, rendering charge carriers and conducting planes, respectively. The $T_c$ in this class of materials generally depends on (i) the size of the RE$^{3+}$ cations, (ii) F substitution on oxygen sites, and (iii) oxygen deficiency in F-free materials. Several reports have shown that $T_c$ can be effectively increased to above 55 K by substitution of larger La$^{3+}$ ion with RE$^+$ cations having smaller radii such as Ce, Pr, Nd, Sm, etc. [2-8]. The pristine (F free) REFeAsO compounds are non-superconducting and show a crystallographic phase transition around 150 K along with a static spin density wave (*SDW*) like long range ordering of the Fe spins around the same temperature. These near concomitant structural and the magnetic transitions are confirmed by an anomaly, which is generally seen as a sharp metallic step at around T$_{SDW}$~150 K in the temperature dependence of resistivity ($\rho$ - $T$) and also by the hump around the same temperature in the heat capacity [1-10]. Introduction of carriers, either by O-deficiency or F-doping is observed to simultaneously shift the magneto-structural transitions to lower temperatures and induce superconductivity.

In the REFeAsO system, the negatively charged Fe-As layers that are sandwiched by positively charged insulating RE-O charge reservoir layers are responsible for the conduction and superconductivity [1-7, 9]. Superconductivity can be induced by (i) F substitution at O sites in RE-O layer [1-9] and (ii) substitution of 3d metals like Co and Ni at Fe sites [11, 12]. The $T_c$ of the 3d metal substituted compounds is generally found to be lower and is believed to be due to the enhanced impurity scattering and pair breaking in the conduction layers. In contrast, the F substitution shows the highest $T_c$ of up to 55 K. The upper critical field ($Hc_2$) is observed to increase appreciably with lowering of the RE cationic size. For example it has been estimated that $Hc_2$ ~ 65 T, ~70 T and ~230 T for $LaFeAsO_{0.9}F_{0.1}$, $PrFeAsO_{0.85}F_{0.15}$, and high-pressure fabricated $NdFeAsO_{0.82}F_{0.18}$, respectively. For $SmFeAsO_{0.85}F_{0.15}$, the value of $H_{c2}$ is well over 200 T [13-16]. Thus Sm substituted compound appears to be potential candidate for high current/magnetic field applications. The tremendous interest in these compounds is not only due to their high critical temperatures, high upper critical fields, and the ability to sustain high current densities at elevated temperatures but also due to the intriguing presence of well-known magnetic iron (Fe) element in the superconducting Fe-As plane. The micro-structural aspects of the Fe-As and RE-O layers and their impact on the superconducting properties are not very well understood yet.

The current article deals with the emergence of superconductivity in F-substituted $SmFeAsO_{0.85}F_{0.15}$ with $T_c$~56 K, which is the highest to the best of our knowledge for bulk samples prepared by normal two-step synthesis method at low temperature and without invoking high *HPHT*. The previous works have reported such high $T_c$ values in excess of ~55K, with *HPHT* synthesis process [8, 17]. We have studied the correlation between the micro-structural properties and superconducting characteristics such as the $T_c$ and the critical current density ($J_c$) along with the flux pinning behaviour of the bulk samples synthesised at low temperature by a two-step solid state reaction route.

**Experimental details**

Polycrystalline bulk samples with nominal composition $SmO_{1-x}F_xFeAs$ (x=0.0 and 0.15) were synthesized by conventional solid state sintering method using high-purity As, Fe, $SmF_3$ and $Sm_2O_3$ powders all with high purity from Sigma Aldrich as starting materials. To obtain SmAs-$Fe_2As$-FeAs powder, Sm, Fe and As were weighed according to the stoichiometric ratio of 1:3:3, mixed and ground thoroughly using mortar and pestle under high purity Ar atmosphere in glove box. The humidity and oxygen content in the glove box was maintained to be less than 1ppm. The mixed powder was pelletized in disk shape and then encapsulated in an evacuated ($10^{-3}$ Torr) silica tube. Then the silica tube embodying the said pellet was heat treated for 12 hrs at 800 °C. Further, in the next step, SmAs, $Fe_2As$ and FeAs were mixed with dehydrated $Sm_2O_3$, Sm and $SmF_3$ in accordance with $(1 + x) Sm + (1 - x) Sm_2O_3 + xSmF_3 + 3FeAs$ stoichiometric ratio, where x = 0 for undoped and 0.15 for fluorine doped sample, was compacted and finally heated at 950°C for 72 hrs in continuum with slow heating rate to obtain a sintered pellet. To prevent silica tube from collapsing during the reaction, the tube was filled with high purity Ar gas. The furnace was then allowed to cool in a natural way. The sintered sample was obtained by breaking the quartz tube. The as sintered sample is concrete ceramic-like with dark-coloured surface.

Phase identification and crystal structure investigation were carried out by using powder *X*-ray diffraction (*XRD*) with Philips *X'*PERT, PRO PAN Analytical X-ray diffractometer with Cu-Kα irradiation (wavelength, $\lambda = 1.5406$ Å) at a scanning speed of $0.02° s^{-1}$. The lattice parameters derived from *X*-ray diffraction patterns were subjected to Rietveld refinement using the FULLPROF SUITE program. Microstructures of the samples under investigations and its morphology were studied using environmental scanning electron microscopy (*SEM*, Quanta 200), operated at 30 kV. Further studies to unravel the microstructure, microstructural characterization were carried out by high resolution transmission electron microscopy (*HRTEM*, *FEI* Tecnai 20G2 operated at 200kV). The resistivity measurements were performed by a conventional four-point-probe method on a Quantum Design Physical Property Measurement System (*PPMS*-140kOe). Magnetic, transport and thermal properties of the samples were carried out with a Physical Property Measurement System (*PPMS*-Quantum Design) in the temperature range 2-300K and in field up to 13Tesla.

**Results and discussion**

The typical *XRD* patterns of the as synthesized pristine $SmFeAsO$ and $SmFeAsO_{0.85}F_{0.15}$ samples are shown in Fig. 1. The Rietveld refinement of X-ray diffraction data collected at room temperature confirmed that all observed reflections are indexed on the basis of tetragonal ZrCuSiAs-type structure with a space group *P*4/*nmm*. The simulated pattern shows that the studied samples are nearly single phase except for a weak impurity diffraction peak (marked with asterisk in the *XRD* pattern) in the F-doped compound. These impurity peaks are assigned to the rare earth oxides. Based on the known structural details Sm and As are taken to be located at the Wyckoff position 2*c* (1/4, 1/4, *z*), the Fe at 2*b* (3/4, 1/4, 1/2), and the O/F are shared at site 2*a* (3/4, 1/4, 0). The lattice parameters obtained from the refinement are $a = 3.937(5)$Å, $c = 8.509(6)$Å and $a = 3.931(6)$Å, $c = 8.474(1)$Å for the pristine and 15% F-substituted samples, respectively. The smaller value of *c*-parameter in the F-substituted sample indicates a successful chemical substitution of $F^{-1}$ ($R_F = 1.33$Å) at $O^{2-}$ ($R_O = 1.40$Å) site. These evaluated values of lattice parameters of the synthesized samples are close to those reported earlier [3, 13-14], which suggests a successful synthesis of envisaged material. The observed structural results are quite consistent with the reports on other rare earth substitutions, and indicate the covalent character of the intra-layer chemical bonding due to the smaller covalent radius of fluorine than oxygen.

To understand the morphology and microstructure of the doped sample and their possible correlation with superconducting properties, *SEM* and *TEM* studies have been carried out on the same sample. Fig. 2(a) shows a representative scanning electron micrograph of SmFeAsO$_{0.85}$F$_{0.15}$ sample. The observation of the fractured specimen clearly shows the occurrence of layered structural units as the constituents of the polycrystalline bulk. This is consistent with the layered crystal structure of the compound. This type of feature has also been observed in high $T_c$ cuprate superconductors. Since the grain boundaries in the samples are assumed to play a crucial role in determining the J$_c$ and as it is strongly influenced by microstructure, extensive local area structural analysis has been done by employing *HRTEM*. In this approach, randomly selected grains have been examined and, during this course; it has been observed that there is an amorphous layer present around most of the grains in this sample. A typical example of this observation is presented in Fig. 2(b), the average thickness of the grain boundary in the present sample is ~ 5nm. The corresponding *SAED* pattern is shown in Fig. 2(c), which confirms the proper phase formation of SmFeAsO$_{0.85}$F$_{0.15}$. Evidence of a similar amorphous layer was also observed in polycrystalline Sr$_{0.6}$K$_{0.4}$Fe$_2$As$_2$ and YBa$_2$Cu$_3$O$_7$ [19, 20]. Generally, this amorphous layer results in the formation of weak link type grain boundaries (*GBs*) which in turn are detrimental to the intergrain current transport. Some reports have also suggested that most of the *GBs* of iron-based superconductors are indeed weak link type [20, 21] and generally resemble to *GBs* in the layered cuprate superconductors. The presence of weak links and the corresponding weakly pinned inter-granular Josephson vortices are responsible for the thermally activated flux flow (*TAFF*) resistivity, which leads to both low $J_c$ and the Arrhenius temperature dependence of resistivity.

Fig. 3(a) shows the temperature dependence of resistivity for SmFeAsO and SmFeAsO$_{0.85}$F$_{0.15}$ samples. The resistivity-temperature ($\rho$ - *T*) curve of undoped SmFeAsO sample exhibits a sudden decrease at $T_{\text{anomaly}}$ ~150K. The anomaly apparent in the resistivity is known to be due to the collective effect of a crystallographic phase transition from the tetragonal *P*4/*nmm* to the orthorhombic *Cmma* space group around *T*~150K, and the occurrence of static spin density wave (*SDW*) instability like magnetic ordering of the Fe spins at a slightly lower temperature of ~140K [5,7,12-18]. The resistivity behaviour of SmFeAsO is semiconducting above 150K and step like metallic at lower temperatures. After substitution of 15% $O^{2-}$ by $F^{1-}$, the resistivity decreases monotonously with decreasing temperature. This decrease in resistivity suggests that the charge carrier density increases and the anomaly completely disappears with a concomitant appearance of superconductivity at below around $T_c$ ($\rho=0$)~55 K. Thus, the $\rho$ - $T$ curve of $SmFeAsO_{0.85}F_{0.15}$ shows the typical metallic behaviour till the superconducting transition is arrived at. The superconducting transition width, $\Delta T_c$ ($T_c^{onset}$ - $T_c^{\rho=0}$), is found to be $\Delta T_c$ ~ 3.1 K. The value of resistivity changes from 15.07mΩ-cm at $T$~300K to 2.7mΩ-cm just above the transition temperature. So the residual resistivity ratio (*RRR*) for our studied $SmFeAsO_{0.85}F_{0.15}$ sample is 5.58, suggests good metallic normal-state connectivity. The broad transition is quite similar to what has been seen in a SmFeAsOF sample synthesized by two-step method. We would like to mention that the observed $T_c$~55K of our F-substituted ($SmFeAsO_{0.85}F_{0.15}$) sample is to the best our knowledge the highest for bulk samples prepared by a low temperature, two-step solid state reaction route and also without employing *HPHT*. $T_c \geq 55K$ has been reported previously but only in materials synthesized through *HPHT* [8, 17]. The normal state resistivity above superconducting transition for the studied $SmFeAsO_{0.85}F_{0.15}$ sample shows linear dependence on temperature. The blue solid line in Fig. 3(a) shows the fitted resistance plot according to equation $\rho = \rho_0 + AT$, where $\rho_0$ is the residual resistivity and A is the slope of the graph. The experimental data fits well in the low temperature range. The linear behaviour of resistivity with temperature deviates above $T$~250K. The values of $\rho_0$ and A are found to be as $2.28 \times 10^{-7}$mΩ-cm and $5.19 \times 10^{-8}$mΩ-cm/K respectively. This behaviour is quite different from those shown by $MgB_2$ [22] and other high-$T_c$ superconductors [23].

To obtain information about upper critical field $Hc_2$ and the flux pinning properties the temperature dependence of electrical resistivity under applied magnetic field [$\rho$ ($T$, $H$)] from 0 to 130kOe for superconducting $SmFeAsO_{0.85}F_{0.15}$ sample in the superconducting range has been measured and shown in Fig. 3(b). The resistive transition at $H=0$ gets significant broadening in the applied magnetic field and is observed to become broader with the increasing $H$. This is generally regarded as a signature of strong vortex motion. This transition broadening with increasing magnetic field $H$, is a characteristic of type-II superconductors, especially the layered high $T_c$ cuprate materials [23]. Further, it is interesting to note that the onset transition temperature ($T_c^{onset}$) remains nearly invariant with respect to $H$, the $T_c$ ($\rho=0$) is observed to decrease from ~ 52K to ~42K at $H$= 130 kOe. This could be due to the granular nature of the studied polycrystalline material, which promotes flux creep behaviour. The lowering of $T_c$ ($\rho$=0) with increasing $H$ may be regarded as the coupled effect of the weal linked nature of the *GB*s and the vortex flow behaviour. The rate of decrease of transition temperature with applied magnetic field $dT_c/dH$ ~0.8 K/T, which is appreciably smaller than the values measured for high $T_c$ superconductor like, YBCO ($dT_c/dH$ ~ 4 K/T, ref. 23) and $MgB_2$ ($dT_c/dH$ ~ 2 K/T, ref.22). This suggests a high value of upper critical field ($H_{c2}$) in these compounds. The upper critical field, $H_{c2}$, defined as the field at which the resistivity increases and approaches to the normal state resistivity, is an important parameter for evaluating the suitability of superconductor for high current/field application. The upper critical field $Hc_2$ has been evaluated by applying the criterion of 90% of normal resistivity at the onset temperature and by using a single-band Werthamer–Helfand–Hohenberg (*WHH*) model [24]. The experimental and the fitted values of the $H_{c2}$ are presented in Fig. 3(c). The slope $dHc_2/dT$ for 90% $\rho_n$ is estimated to be -13.93T/K. Similarly the slope $dH_{c2}/dT$ for 10% $\rho_n$ is -1.89 T/K at $T \leq 52$ K. The relation between $dH_{c2}/dT$ and $H_{c2}(0)$ is defined by Werthamer–Helfand–Hohenberg (*WHH*) model defined by:

$$\qquad\qquad$$

We estimated the $H_{c2}$ values for the superconducting F-doped compound using the above formula. The estimated value of the upper critical fields are $H_{c2}(0)$ ~536 T and ~68 T for the 90 % and 10 % $\rho_n$ fields, respectively and $H_{irr}$ ~68.5 T by the aforementioned model. The Ginzburg–Landau equation

$$\qquad\qquad$$

where t=T/$T_c$ is the reduced temperature, was used for the fitting and extrapolation of the $H_{c2}$-T data. The experiment and fitted data is shown in Fig. 3c. The above equation gives $H_{c2}$ ~565.80 T, which is slightly higher than that estimated by *WHH* approach. These estimated values of $H_{c2}$ clearly demonstrate the robustness of the present ~55 K superconducting $SmFeAsO_{0.85}F_{0.15}$ against the applied magnetic field. The above estimations are only suitable for the one-band case. However, in this new type of superconductor, the band structure calculation indicates the existence of multiple bands across the Fermi levels. As our analysis neglects multiple band effect, the real upper critical fields in the present system could be very high upper critical field suggests a very prospective application of this superconducting envisaged material. Assuming $H_{c2}(0) = H_{c2}(WHH_0)$, the Ginzburg–Landau coherence length $\xi_{GL}= (\Phi_0/2\pi H_{c2})^{1/2}$, where $\Phi_0=2.07\times10^{-7}$G cm$^2$ is the flux quantum, yields zero temperature coherence length $\xi_{GL}(0) \sim 7.8$ Å. This value is smaller than the periodicity of FeAs layer of d=8.5 Å. This clearly suggest a two dimensional behavior in the lower temperature regime.

The temperature derivative of resistivity for the superconducting SmFeAsO$_{0.85}$F$_{0.15}$ sample at various applied fields has been shown in inset of Fig. 3 (d). The d$\rho$/d$T$ shows a narrow intense maxima centred at superconducting transition temperature in zero applied fields, which indicate good percolation path of superconducting grains. The broadening of the d$\rho$/d$T$ peak increases with applied magnetic field. The broad peak at low temperatures range indicates the intra-grain and inter-grain regimes [23]. It is interesting to note that a clear second peak in d$\rho$/d$T$, usually observed in case of the high $T_c$ cuprate materials [23], is absent in the presently studied SmFeAsO$_{0.85}$F$_{0.15}$. This suggests that unlike the insulating *GB*s between superconducting grains in the high $T_c$ cuprate materials, the same are mostly metallic in the case of SmFeAsO$_{0.85}$F$_{0.15}$. On the other hand, the broadening in d$\rho$/d$T$ peak with applied field, suggestive of weaker inter granular coupling, is relatively higher for SmFeAsO$_{0.85}$F$_{0.15}$ than for MgB$_2$ [22].

The broadening of resistive transition in applied magnetic field is due to the creep of vortices and as a result the dependence is thermally activated. The temperature dependence of resistivity in the broadened region is generally described by Arrhenius equation $\rho(T,B) = \rho_0 \exp[-U_0/k_BT]$, $\rho_0$ is the field independent pre-exponential factor (here normal state resistance at 55K($\rho_{55}$) is taken as $\rho_0$) and k$_B$ is Boltzmann's constant and U$_0$, the flux-flow activation energy which can be obtained from the slope of the linear part of an Arrhenius plot[25,26]. The Arrhenius plot of $\rho(T)$ shows linearity in limited temperature interval below $T_c$, the activation energy U$_0$ can be deduced from this region. The Arrhenius plots of $\rho(T)$ for superconducting SmFeAsO$_{0.85}$F$_{0.15}$ sample in applied fields of up to 130kOe are presented in Fig. 4. The best fitted experimental data gives the value of activation energy (U$_0$/k$_B$) ranging from 2961K and 662K in fields of 0.1kOe and 130kOe, respectively. The activation energy shows different power law dependence on applied magnetic field [27]. In the present study, the activation energy shows weak field dependence i.e. U$_0$/k$_B \sim H^{-0.06}$ at small magnetic field ($H < 10$ kOe). However, at higher values of $H$, the field dependence is much stronger; U$_0$/k$_B \sim H^{-0.66}$. Inset of Fig. 4 depicts the magnetic field dependence (up to 130kOe) of the activation energy U$_0$/k$_B$ for SmFeAsO$_{0.85}$F$_{0.15}$ sample. These values are in good agreement with previous reports [12, 27].

To understand the magnetic behaviour comprehensively the temperature and field dependent magnetization measurements have been carried out. Fig. 5(a) shows the temperature dependence of the *DC* magnetization of the SmFeAsO$_{0.85}$F$_{0.15}$ sample measured under zero field cooled (*ZFC*) and field cooled (*FC*) measuring conditions at *H*=10 Oe. The negative susceptibility in both *ZFC* and *FC* condition confirm the bulk superconductivity. The susceptibility signal becomes negative below 54K. This indicates that bulk superconductivity sets below this temperature. Both *ZFC* and *FC* curves are nearly saturated below 45K. The minute difference in $T_c$ (onset) obtained from resistivity *ρ(T)* and magnetization *M(T)* is due to the threshold for transport measurements. The transport measurement is through percolation path and hence is lower in comparison to bulk diamagnetic response. The difference in *ZFC* and *FC* curves of the sample also suggests that the material has a fairly large flux pinning force resulting in the trapping of magnetic flux under the field cooling condition. Fig. 5(b) depicts the isothermal magnetization *M(H)* loops of the superconducting SmFeAsO$_{0.85}$F$_{0.15}$ sample at 5K, 10K and 200K, with applied fields of up to 20kOe. The *M* (*H*) loops are wide open at 5K and 10K up to 20kOe and thus confirming the bulk superconductivity in the studied sample. The lower critical field ($H_{c1}$) i.e. the characteristic value at which the applied magnetic field starts to penetrate the sample has been deducted from the *M(H)* loops. The magnetic field, at which the magnetization deviates from the linearity, is also defined as lower critical field of superconducting materials. The variation of Magnetization under applied magnetic field, *M* (*H*) plots for the SmFeAsO$_{0.85}$F$_{0.15}$ sample at temperature 5K and 10K are shown in the Fig. 5(c). The lower critical field deducted from the graph is found to be $H_{c1}$~ 650Oe at 5K and 10K.

Fig. 6 shows the magnetic field dependence of the critical current density $J_c$ being derived from the hysteresis loop (*MH*) width using the extended Bean critical state model $J_c$ = 20Δ*M*/Va(1-a/3b), where Δ*M* is the height of the magnetization loop measured in emu, V is the volume of the sample in cm$^3$, a and b are the respective sample dimensions in cm and $J_c$ is in Acm$^{-2}$. For estimation of $J_c$, the full sample dimensions of 7.1 × 2.62 × 2.28mm$^3$ were taken. At 5K, the $J_c$ is found to be approximately 5.26 × 10$^4$ cm$^{-2}$ at *H*=0 which decreases to 2.2 × 10$^3$Acm$^{-2}$ at *H*=100 kOe. It can also be observed from Fig. 6 that $J_c$ at 5K decreases rapidly up to *H*= 20kOe. These values of J$_c$ are found to be in good agreement with the reported results for the SmFeAsO$_{0.85}$F$_{0.15}$ superconductor [28]. However, $J_c$ values of SmFeAsO$_{0.85}$F$_{0.15}$ superconductor at 5K obtained is significantly lower than that of the MgB$_2$ bulk samples which generally attain 10$^6$Acm$^{-2}$ at 4.2K. The possible reasons for lower value of $J_c$ could be the presence of second minor impurity phase or the weak link nature of the *GB*s. Although the whole-sample current densities are significantly lower than that in randomly grain-oriented bulk of $MgB_2$, the grain connectivity seems to be better than that of random polycrystalline high $T_c$ cuprates.

In order to understand granular nature of the superconducting sample and in particular to find out the intra and inter grain contributions at high temperatures, the ac susceptibility measurement for the superconducting sample at zero *DC* bias magnetic fields has been performed [19, 29, 30]. The real ($\chi'$) and imaginary ($i\chi''$) components of ac susceptibility as function of temperature measured at various *AC* drive field, ranging from 3Oe to 13Oe at fixed frequency f = 333Hz are presented in Fig. 7. It is clear from the real part of the susceptibility ($\chi'$) that the diamagnetic onset transition temperature is same irrespective to the drive fields and in imaginary ($i\chi'$) components of ac susceptibility two peaks appear in which the higher temperature peak corresponds to the individual superconducting grains i.e. intra granular superconductivity whereas the lower temperatures peak corresponds to the intergranular coupling. The intergranular peak shifts towards lower temperature with increasing ac drive fields and gets broadened, due to the weak coupling between the grains [31].

The temperature dependence of specific heat for the pure and doped samples is shown in Fig. 8. The absolute values of $C_p$ are quite close for both the samples. A clear jump is observed in $C_p$ data of SmFeAsO around the same temperature at which a metallic step has been seen (Fig. 2(a)) in resistivity measurements. This jump in SmFeAsO heat capacity around 140K is due to the spin density wave (*SDW*) character exhibited by the compound. Such jump in specific heat completely disappears in superconducting sample. Heat capacity decreases with further decrease in temperature and an additional peak is observed at 4.5K which may be due to the antiferromagnetic ordering of $Sm^{3+}$ ions. The lower inset of Fig. 7 shows measured value of $C_p/T$ Vs $T^2$. At the superconducting transition temperature ($T_c$), a hump in the $Cp/T$ Vs $T^2$ plot has been observed. The shape of the Lambda transition for superconducting $SmFeAsO_{0.85}F_{0.15}$ does not exhibit sharp discontinuity at $T_c$, as has been reported for other superconductors [32, 33]. In general, the heat capacity of non superconducting SmFeAsO and superconducting $SmFeAsO_{0.85}F_{0.15}$ exhibits clearly the coupled *SDW*/structural phase transition at around 150K for the former and a discontinuity in the $Cp/T$ vs $T^2$ for the later at superconducting transition temperature of 55K.

To elucidate the non-magnetic contribution of heat capacity and change in entropy for anti-ferromagntic ($T_N$) ordering of $Sm^{3+}$ spins in non superconducting SmFeAsO and superconducting $SmFeAsO_{0.85}F_{0.15}$ samples, the polynomial interpolation method using equation $aT+bT^3$ has been used to fit the $Cp(T)$ plot of both samples [10,34]. The fitted and observed $C_P(T)$ are close to the $T_N$ of Sm is shown in Fig. 8(b). Using the fitted value of the coefficients a and b, the data has been extrapolated to estimate the background contribution of heat capacity in the lower temperature range. In order to calculate the change in entropy, the change in specific heat ($\Delta C_P$) is found by first subtracting the background contribution from the experimental data and further the integration over $\Delta C_P/T$ provides the entropy change $\Delta S$. The integrated result ($\Delta S$) provides the values ~4.256J/mole-K and 2.455J/mole-K for SmFeAsO and SmFeAsO$_{0.85}$F$_{0.15}$ superconducting samples, respectively.. The change in entropy near $T_N$ of Sm for both the samples is shown in inset of Fig. 8(b). It is interesting to note that change in entropy ($\Delta S$) for $T_N$ of Sm is reduced by over 40% for the superconducting (SmFeAsO$_{0.85}$F$_{0.15}$) sample in comparison to its non-superconducting (SmFeAsO) counter part.

In summary, the structural, microstructural and magnetotransport properties of SmFeAsO$_{0.8}$5F$_{0.15}$ superconductor synthesized at low temperature by two-step low temperature solid state reaction process have been investigated. In this nearly single phase material, having weak link type *GB*s, the superconducting transition onsets at ~55K and the zero field transition width is ~3K. The *DC* magnetization and *AC* susceptibility measurement also provide clear evidence of weak link nature of the *GB*s. Our results show that despite the weak link nature of the *GB*s the decrement in $T_c$ ($\rho=0$) with magnetic field, $dT_c/dH$~0.8K/T is appreciably smaller than the high $T_c$ cuprates. The observed broadening of resistive transition in applied magnetic field is caused by the creep of flux vortices and motion is thermally activated. At lower *H*, the activation energy shows weak field dependence, while at higher values it becomes much stronger. Although the $J_c$ of our samples is moderate but the decay in it under *H* is comparatively smaller. We believe that more studies are required to further understand the nature of the *GB*s and its consequences on the magneto-transport properties.

**Acknowledgement**

The authors are thankful to Prof. Ratnamala Chatterjee (*IITD*). One of the authors (A.S.) would like to acknowledge *UGC*-India and Dr. U.P. Singh (Principal *TDPG* College) for providing Teacher fellowship award under *FIP*. The work at *CSIR-NPL* is financially supported by *DAE-SRC* outstanding investigator award scheme to work on search for new superconductors.

**Reference**

1. Y. Kamihara, T. Watanabe, M. Hirano and H. Hosono: J. Am. Chem. Soc. 130 (2008) 3296.
2. X.H. Chen, T. Wu, G. Wu, R.H. Liu, H. Chen and D.F. Fang: Nature 453 (2008) 761.
3. Z.A. Ren, J. Yang, W. Lu, W. Yi, X.L. Shen, Z.C. Li, G.C. Che, X.L. Dong, L.L. Sun, F. Zhou and Z.X. Zhao: Europhys. Lett. 82 (2008) 57002.
4. Z.A. Ren, J. Yang, W. Lu, W. Yi, G.C. Che, X.L. Dong, L.L. Sun and Z.X. Zhao: Mater. Res. Innov. 12 (2008) 1.
5. G.F. Chen, Z. Li, D. Wu, G. Li, W.Z. Hu, J. Dong, P. Zheng, J.L. Luo and N.L. Wang: Phys. Rev. Lett. 100 (2008) 247002.
6. C. Wang, L.J. Li, S. Chi, Z.W. Zhu, Z. Ren, Y.K. Li, Y.T. Wang, X. Lin, Y.K. Luo, S. Jiang, X.F. Xu, G.H. Cao and Z.A. Xu: Europhys. Lett. 83 (2008) 67006.
7. L.J. Li, Y.K. Li, Z. Ren, Y.K. Luo, X. Lin, M. He, Q. Tao, Z.W. Zhu, G.H. Cao and Z.A. Xu: Phys. Rev. B 78 (2008) 132506.
8. P. Cheng, L. Fang, H. Yang, X. Zhu, G. Mu, H. Luo, Z. Wang and H.H. Wen: Sci. China G 51 (2008) 719.
9. V.P.S. Awana, R.S. Meena, A. Pal, A. Vajpayee, K.V.R. Rao and H. Kishan: Eur. Phys. J. B 79, (2011) 139.
10. V. P. S. Awana, Anand Pal, Arpita Vajpayee, H. Kishan, G.A. Alvarez, K. Yamaura and E. Takayama-Muromachi: J. Appl. Phys. 105, (2009) 07E316
11. V.P.S. Awana, Arpita Vajpayee, Anand Pal, Monika Mudgel, R.S. Meena and H. Kishan: J Supercond Nov Magn 22, (2009) 623.
12. Anand Pal, S.S. Mehdi, Mushahid Husain and V.P.S. Awana: Solid State Sciences 15 (2013) 123.
13. A. Narduzzo, M. S. Grbić, M. Požek, A. Dulčić, D. Paar, A. Kondrat, C. Hess, I. Hellmann, R. Klingeler, J. Werner, A. Köhler, G. Behr, and B. Büchner: *Phys. Rev. B 78 (2008)012507*.
14. F. Hunte, J. Jaroszynski, A. Gurevich, D. C. Larbalestier, A. S. Jin R, Sefat, M.A. McGuire, B.C. Sales, D. K.Christen and D. Mandrus :Nature *453 (2008) 903*.
15. C. Senatore, R. Flukiger, M. Cantoni, G. Wu, R. H. Liu and X. H. Chen: Phys. Rev. B 78(2008) 054514.
16. X. L. Wang, S. R. Ghorbani, G. Peleckis and S. X. Dou: Adv. Mater. 21 (2008)236.
17. R. Zhi-An, L. Wei, Y. Jie, Y. Wei, S.X. Li, Z. Cai, C. G. Can, D. X. Li, S. L. Ling, Z. Fang and Z. Z. Xian: Chinese Phys. Lett. 25, (2008)2215.
18. M. Tropeano, C. Fanciulli, C. Ferdeghini, D. Marrè, A. S. Siri, M. Putti, A. Martinelli, M. Ferretti, A. Palenzona, M. R. Cimberle, C. Mirri, S. Lupi, R. Sopracase, P. Calvani and A. Perucchi: Super. Sci. & Technol. 25, (2009)034004.
19. L. Wang, Y. Qi, D. Wang, X. Zhang and Y. Ma: Superconductor Sci. and Tech.23, (2010)025027

20. Y. Matsumoto, J. Hombo and Y. Yamaguchi: Appl. Phys. Lett. 56, (1990)1585
21. J.H. Durell, C.-B Eom, A. Gurvich, E.E. Hellstrom, C. Tarantini and D.C. Larbalestier: Rep. Prog. Phys. 74, (2011) 124511.
22. V. P. S. Awana, Arpita Vajpayee, Monika Mudgel, V. Ganesan, A.M. Awasthi, G.L. Bhalla, and H. Kishan: Eur. Phys. J. B 62, (2008)281-294.
23. N. P. Liyanawaduge, S. K. Singh, A. Kumar, R. Jha, B. S. B. Karunarathne, and V. P. S. Awana: Supercond. Sci. Technol. **25**, (2012)035017.
24. E. Helfand and N.R. Werthamer: Phys. Rev. 147, (1966) 288.
25. A. Gurevich: Rep. Prog. Phys. 74 (2011) 124501.
26. J. Jaroszynski J, F. Hunte, L. Balicas, Jo Youn-jung , I. Rai Cevic, A. Gurevich, D. C. Larbalestier, F.F. Balakirev, L. Fang, P. Cheng, Y. Jia, and H. H. Wen: Phys. Rev. B 78 (2008) 174523.
27. C. Shekhar, Amit Srivastava, Pramod Kumar, Pankaj Srivastava and O. N. Srivastava: Supercond. Sci. Technol. 25 (2012) 045004.
28. L. Wang, Z. Gao, Y. Qi, X. Zhang, D. Wang and Y. Ma: Supercond. Sci. Technol. 22, (2009)015019.
29. M. Nikolo and R.B. Goldfarb: Phys. Rev. B 39, (1989) 6615.
30. G. Bonsignore, A. Agliolo Gallitto, M. Li Vigni, J. L. Luo, G.F. Chen, N. L. Wang and D.V. Shovkun: J. Low. Temp. Phys. 162, (2011)40-51.
31. A. Agliolo Gallitto, G. Bonsignore, M. Bonura and M.L Vigni: J. Phys. Conf. Series 234, (2010) 012001.
32. L. Ding, C. He, J. K. Dong, T. Wu, R. H. Liu, X. H. Chen, and S. Y. Li: Phys. Rev. B 77, (2008) 180510(R).
33. M. Tropeano, A. Martinelli, A. Palenzona, E. Bellingeri, E. Galleani d'Agliano, T. D. Nguyen, M. Affronte and M. Putti: Phys. Rev. B 78, (2008) 094518.
34. V.P.S. Awana, Anand Pal, M. Husain · H. Kishan, J Supercond Nov Magn 24, (2011) 151.

**Figure Captions**

**Fig. 1** Rietveld fitted room temperature *X*-ray diffraction patterns of SmFeAsO and SmFeAsO$_{0.85}$F$_{0.15}$. All the permitted diffraction planes are marked with blue vertical lines. The green line in the bottom shows the difference between the observed, Y$^{obs}$ (red open circle) and fitted, Y$^{calc}$ (black line) patterns (Y$^{obs}$- Y$^{calc}$).

**Fig. 2(a)** *SEM*, **(b)** *TEM* and **(c)** *ED* micrographs of studied 55K superconducting SmFeAsO$_{0.85}$F$_{0.15}$ sample

**Fig. 3(a)** Temperature dependence of the resistivity $\rho(T)$ of SmFeAsO and SmFeAsO$_{0.85}$F$_{0.15}$ samples. Black arrow shows the anomaly temperature ($T_m$) for SmFeAsO and red arrow shows the superconducting transition temperature of doped sample. Solid blue line shows the linear fitted data in normal state for SmFeAsO$_{0.85}$F$_{0.15}$.

**Fig. 3(b)** Resistivity behaviour under applied magnetic field $\rho(T)H$, up to 13Tesla for SmFeAsO$_{0.85}$F$_{0.15}$. Inset shows the derivates of resistivity for SmFeAsO$_{0.85}$F$_{0.15}$ in transition region.

**Fig. 3(c)** Dependence of Upper critical field $H_{c2}(T)$ with temperature for 90% and 10 % drop of resistance of the normal state resistance and at onset (T$_c$ $^{onset}$ ).The inset shows estimation of $H_{c2}(T)$ for 90% drop of resistance of the normal state resistance and at onset (T$_c$ $^{onset}$ ) using the G-L approach.

**Fig. 4** The Arrhenius plot of resistivity at different field to study the thermally activation flux flow for smFeAsO$_{0.85}$F$_{0.15}$ sample. Inset show the magnetic field dependence of *TAFF* activation energy ($U_0$).

**Fig. 5(a)** Temperature dependence *DC* magnetic susceptibility *M*(*T*) in both *ZFC* and *FC* situation of SmFeAsO$_{0.85}$F$_{0.15}$ sample under applied field of 10Oe.

**Fig. 5(b)** Variation of Magnetization under applied magnetic field, *M*(*H*) at 5K, 10K and 200K of SmFeAsO$_{0.85}$F$_{0.15}$ sample.

**Fig. 5(c) T**he Lower critical field ($H_{c1}$) at 5 and 10K for the superconducting SmFeAsO$_{0.85}$F$_{0.15}$ sample

**Fig. 6** The Variation of Critical current density ($J_c$) with applied magnetic field.

**Fig. 7** The real ($\chi'$) and imaginary ($i\chi''$) components of *AC* susceptibility as function of temperature measured at various *AC* drive field, ranging from 3Oe to 13Oe at 333*Hz* for SmFeAsO$_{0.85}$F$_{0.15}$ sample.

**Fig. 8(a)** Specific heat of SmFeAsO and SmFeAsO$_{0.85}$F$_{0.15}$ sample (main panel). Inset shows enlarge view of the superconducting transition $C_p/T$ Vs $T^2$ near $T_c$ region.

**Fig. 8(b)** Observed and fitted Specific heat of SmFeAsO near the *AFM* ordering of Sm spins, the inset shows that change in entropy near *AFM* ordering of Sm spins for both non superconducting (SmFeAsO) and superconducting ( SmFeAsO$_{0.85}$F$_{0.15}$) samples.

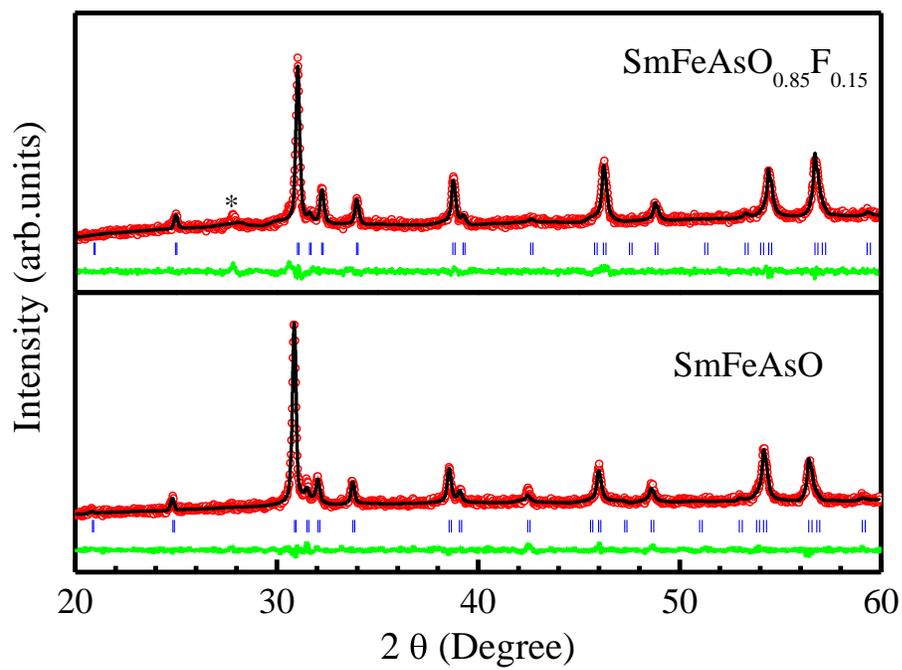

Fig. 1

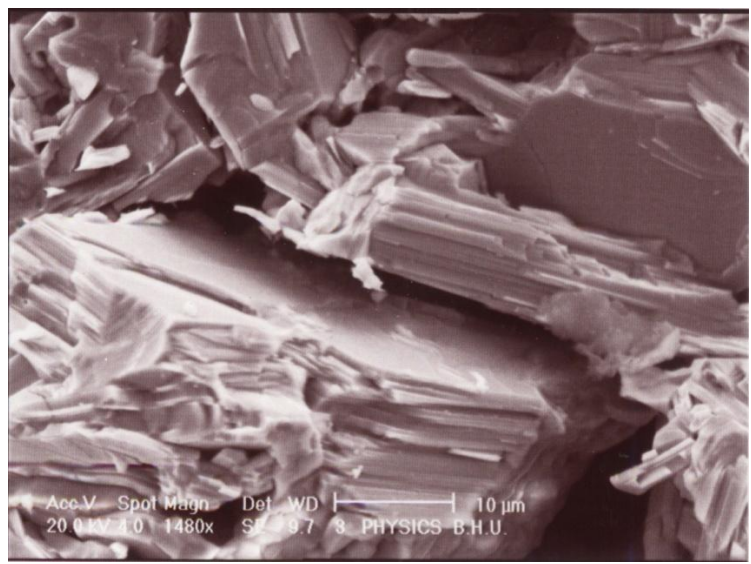

Fig. 2(a)

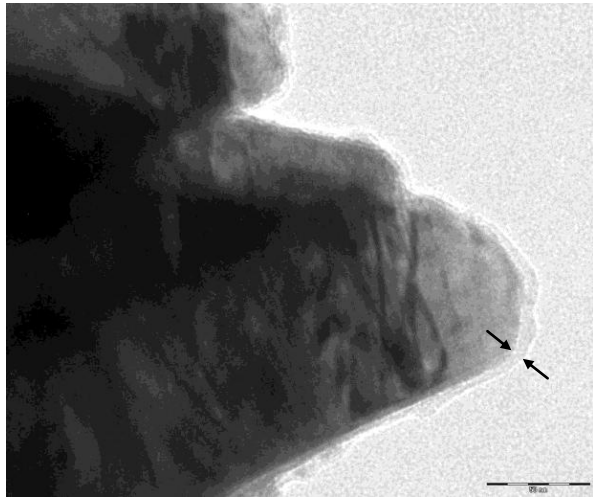

Fig. 2(b)

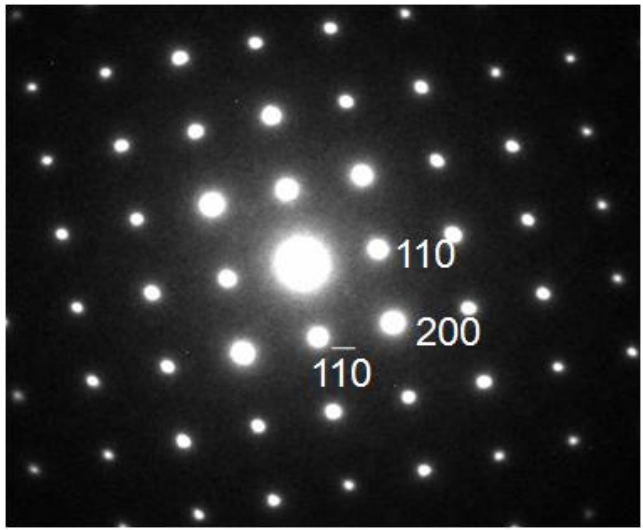

Fig. 2(c)

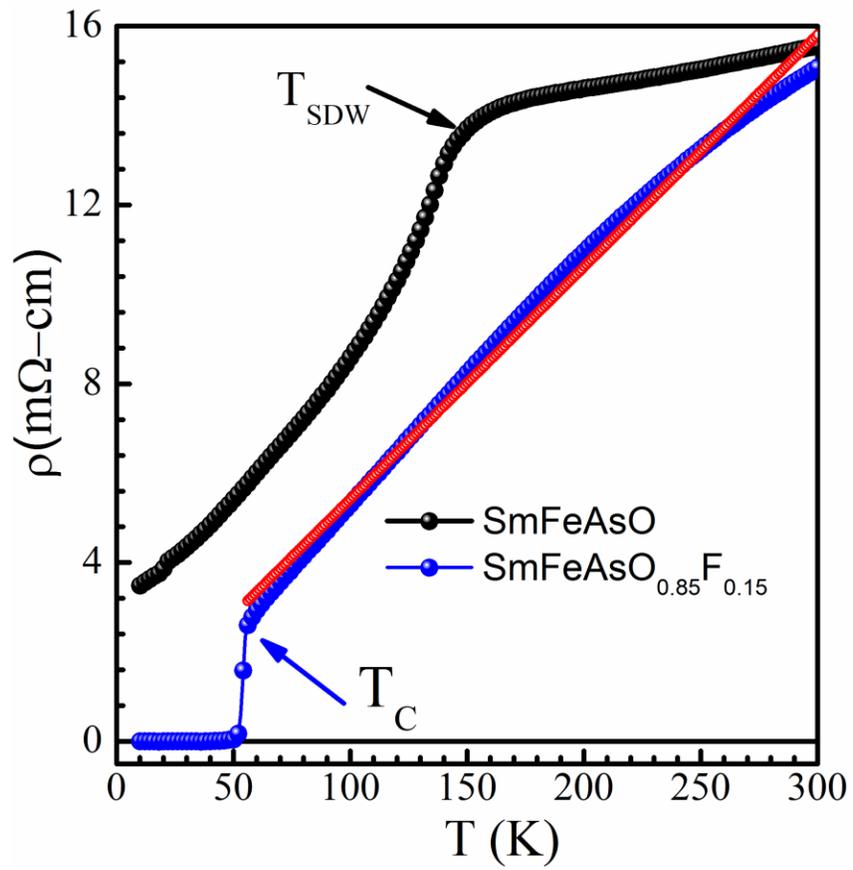

Fig. 3 (a)

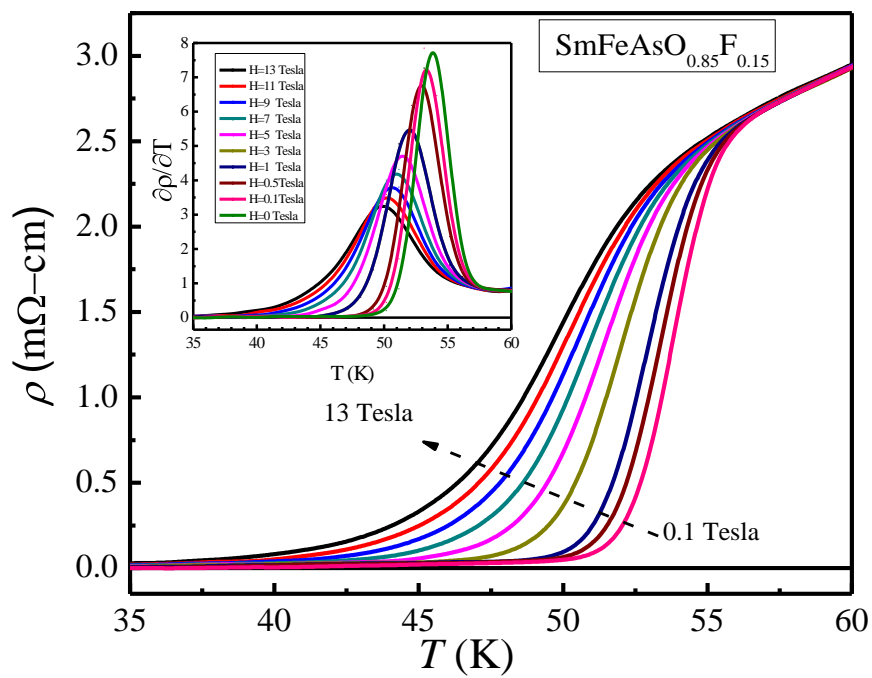

Fig. 3 (b)

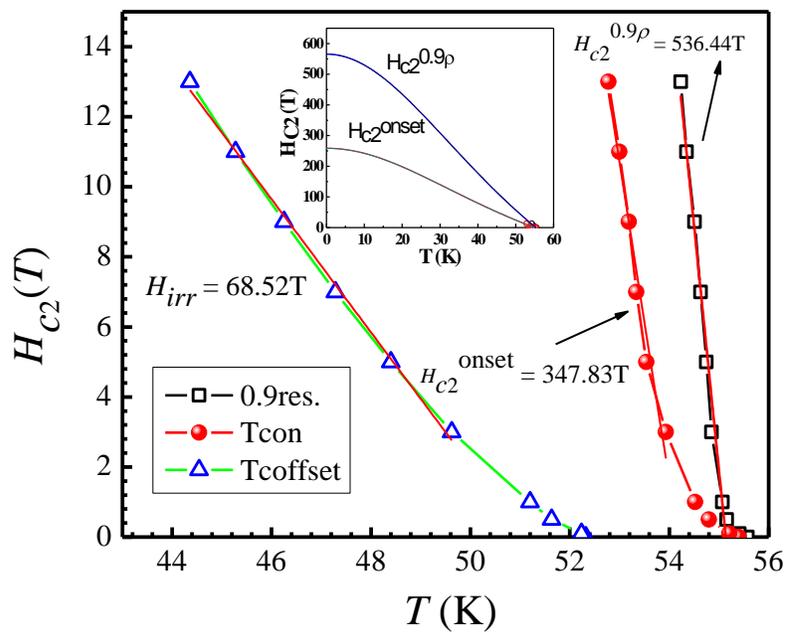

Fig. 3(c)

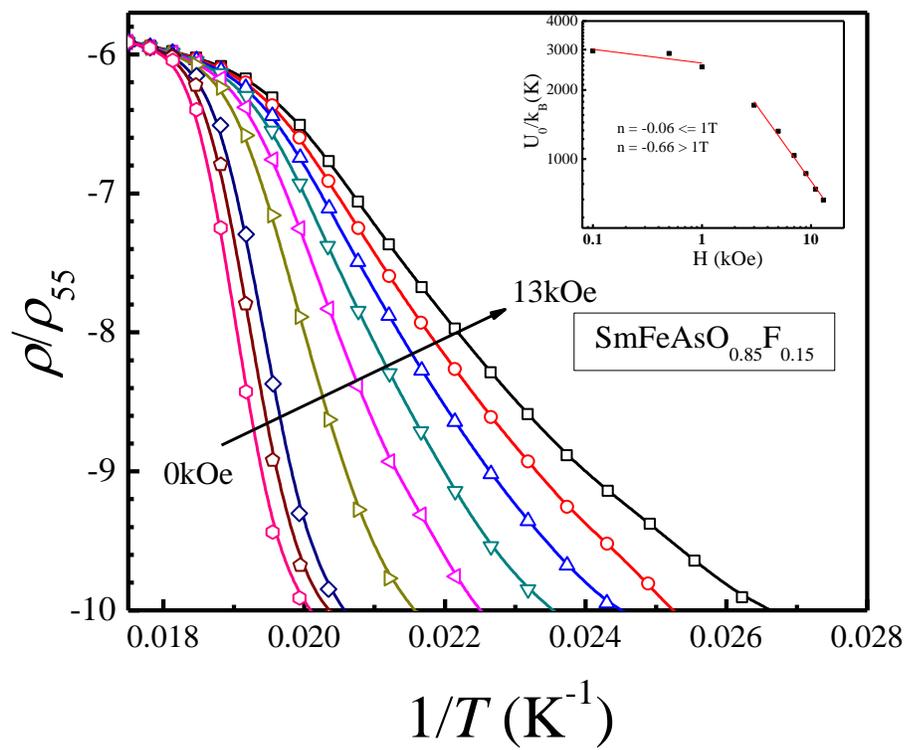

Fig. 4

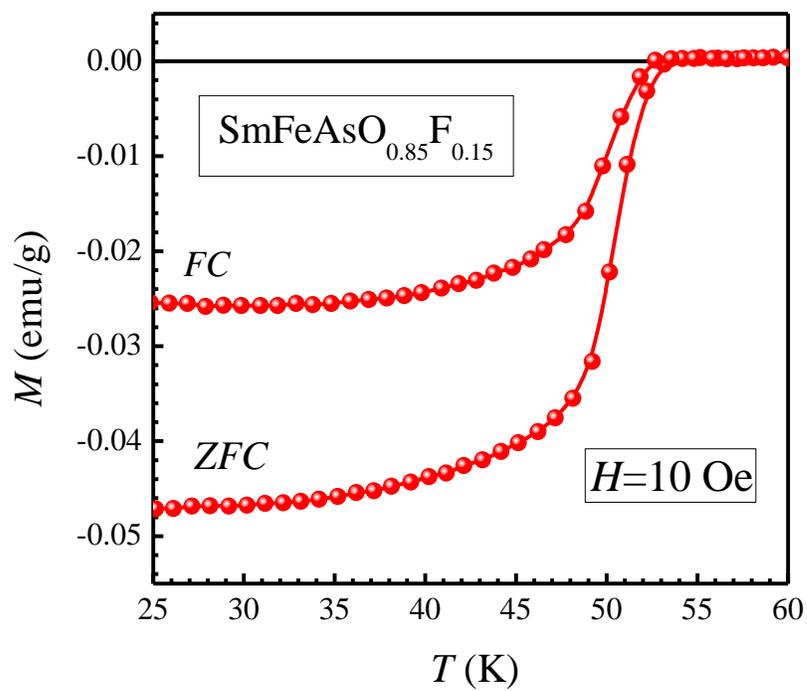

Fig. 5(a)

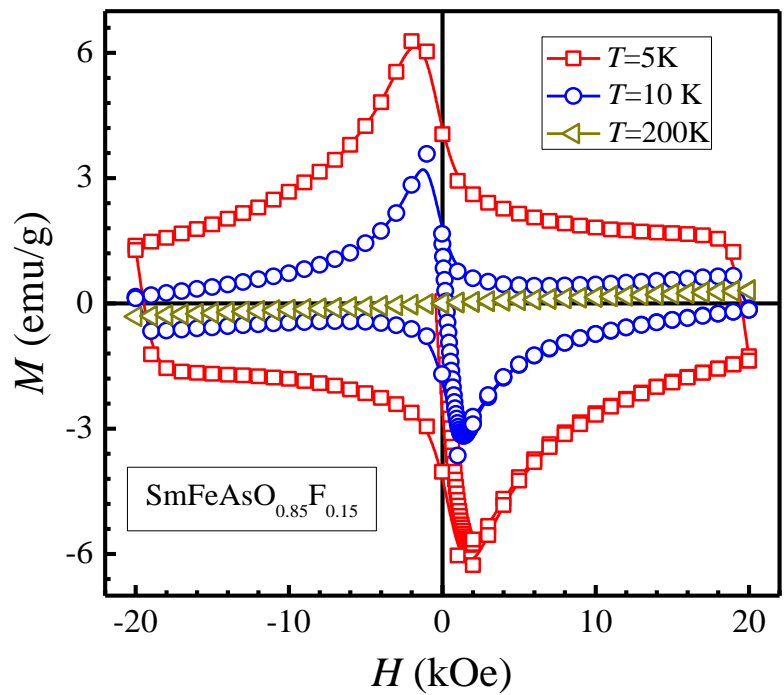

Fig. 5 (b)

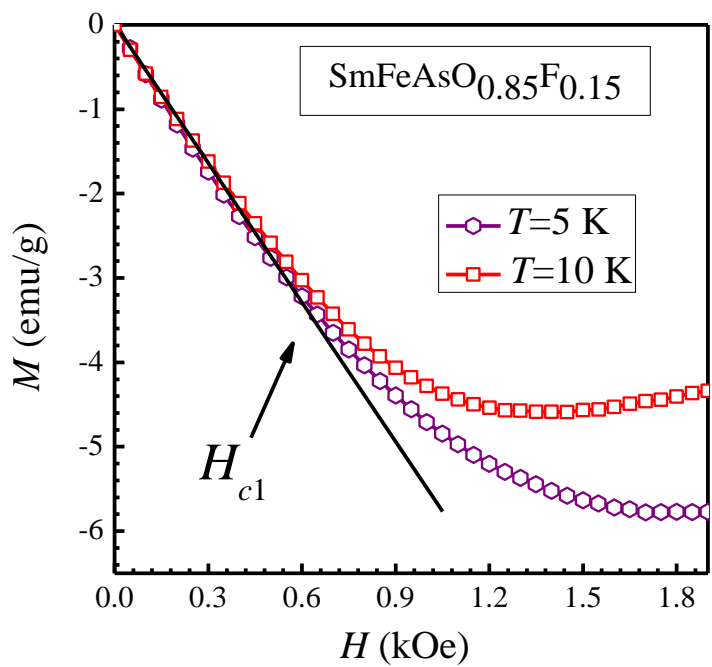

Fig. 5(c)

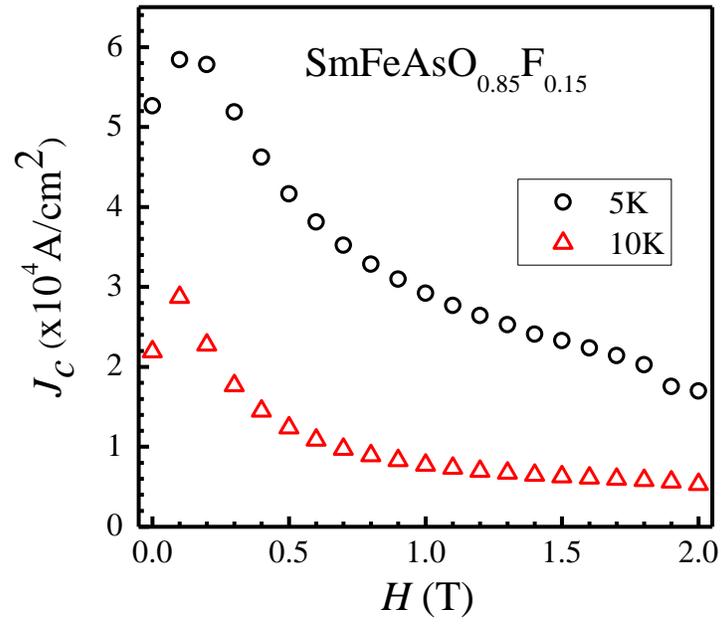

Fig. 6

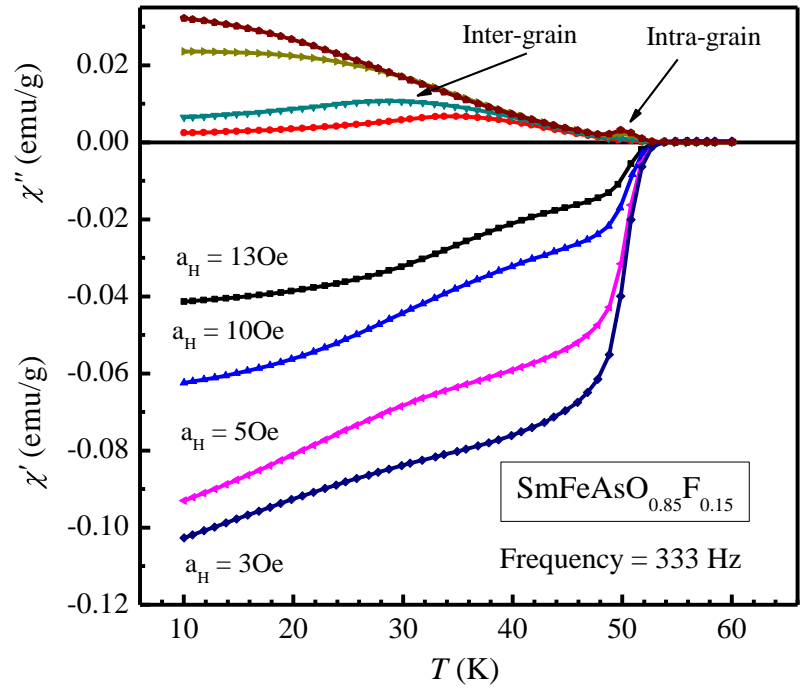

Fig. 7

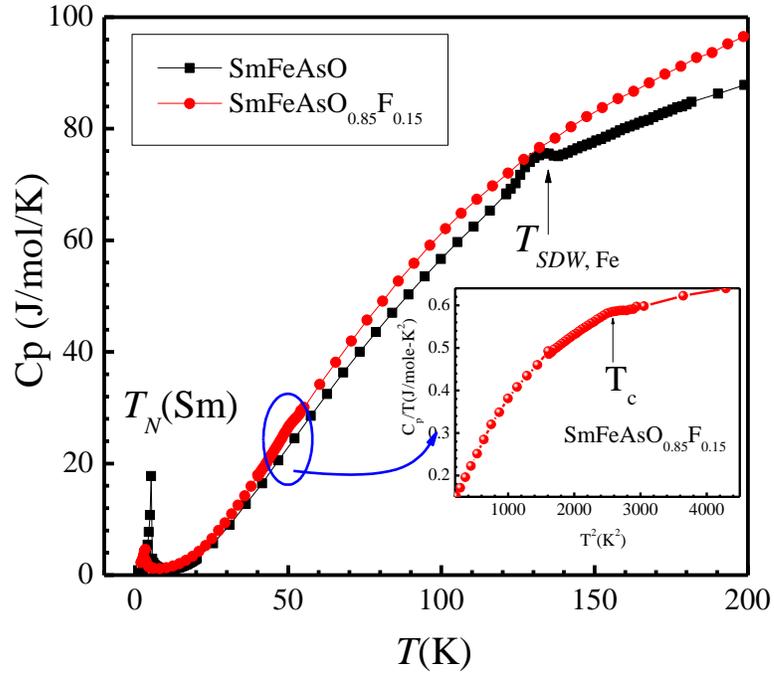

Fig. 8(a)

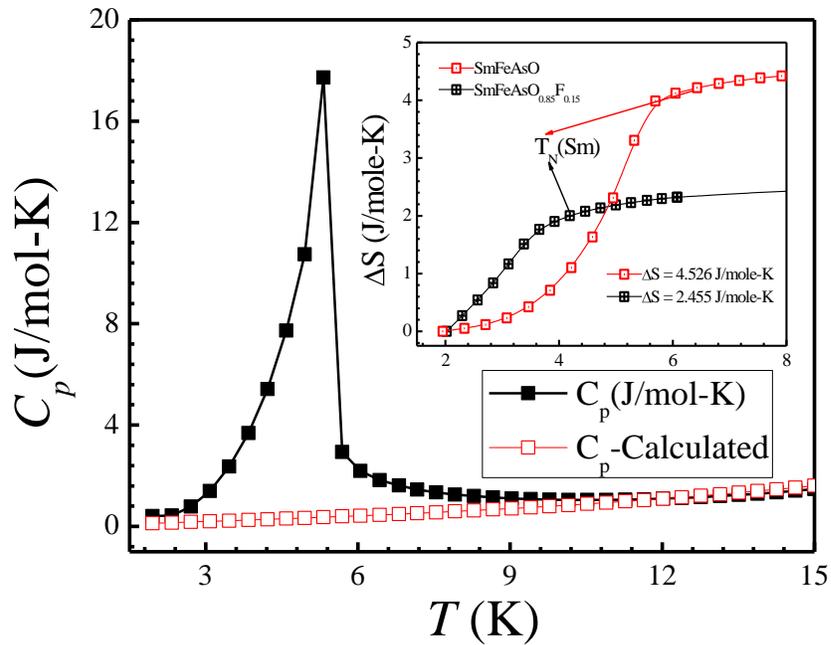

Fig. 8 (b)